\documentclass[preprint,12pt]{elsarticle}

\usepackage{amssymb}
\usepackage{comment}
\usepackage{xr}
\makeatletter
\newcommand*{\addFileDependency}[1]{% argument=file name and extension
  \typeout{(#1)}
  \@addtofilelist{#1}
  \IfFileExists{#1}{}{\typeout{No file #1.}}
}
\makeatother

\newcommand*{\myexternaldocument}[1]{%
    \externaldocument{#1}%
    \addFileDependency{#1.tex}%
    \addFileDependency{#1.aux}%
}
\myexternaldocument{supplementary}
\journal{Journal of Colloid and Interface Science}

\begin{document}

\begin{frontmatter}

%% Title, authors and addresses

%% use the tnoteref command within \title for footnotes;
%% use the tnotetext command for theassociated footnote;
%% use the fnref command within \author or \address for footnotes;
%% use the fntext command for theassociated footnote;
%% use the corref command within \author for corresponding author footnotes;
%% use the cortext command for theassociated footnote;
%% use the ead command for the email address,
%% and the form \ead[url] for the home page:
%% \title{Title\tnoteref{label1}}
%% \tnotetext[label1]{}
%% \author{Name\corref{cor1}\fnref{label2}}
%% \ead{email address}
%% \ead[url]{home page}
%% \fntext[label2]{}
%% \cortext[cor1]{}
%% \affiliation{organization={},
%%             addressline={},
%%             city={},
%%             postcode={},
%%             state={},
%%             country={}}
%% \fntext[label3]{}

\title{Probing surfactant bilayer interactions by tracking optically trapped single nanoparticles}

%% use optional labels to link authors explicitly to addresses:
%% \author[label1,label2]{}
%% \affiliation[label1]{organization={},
%%             addressline={},
%%             city={},
%%             postcode={},
%%             state={},
%%             country={}}
%%
%% \affiliation[label2]{organization={},
%%             addressline={},
%%             city={},
%%             postcode={},
%%             state={},
%%             country={}}

\author[NAM]{Jeonghyeon Kim}
\ead{jeonghyeon.kim@epfl.ch}
\author[NAM]{Olivier J. F. Martin\corref{corr}}
\cortext[corr]{Corresponding author}
\ead{olivier.martin@epfl.ch}
\affiliation[NAM]{organization={Nanophotonics and Metrology Laboratory, Swiss Federal Institute of Technology Lausanne (EPFL)},%Department and Organization
%            addressline={Address One}, 
            city={Lausanne},
            postcode={1015}, 
%            state={State One},
            country={Switzerland}}

\begin{abstract}
%% Text of abstract
Single-particle tracking and optical tweezers are powerful techniques for studying diverse processes at the microscopic scale. The stochastic behavior of a microscopically observable particle contains information about its interaction with surrounding molecules, and an optical tweezer can further facilitate this observation with its ability to constrain the particle to an area of interest. Although these techniques found their initial applications in biology, they can also shed new light on colloid and interface phenomena by unveiling nanoscale morphologies and molecular-level interactions in real time, which have been obscured in traditional ensemble analysis. Here we demonstrate the application of single-particle tracking and optical tweezers for studying molecular interactions at solid--liquid interfaces. Specifically, we investigate the behavior of surfactants at the water--glass interface by tracing their interactions with gold nanoparticles that are optically trapped on these molecules. We discover the underlying mechanisms governing the particle motion, which can be explained by hydrophobic interactions, disruptions, and rearrangements among surfactant monomers at the interfaces. Such interpretations are further supported by statistical analysis of an individual trajectory and comparison to theoretical predictions. Our findings provide new insights into the surfactant dynamics in this specific system but also illustrate the promise of single-particle tracking and optical manipulation in studying nanoscale physics and chemistry of surfaces and interfaces. 

\end{abstract}

\end{frontmatter}

%% \linenumbers

%% main text
\section{Introduction}
% 1. Context
% define Single particle tracking
% the usage of single particle tracking
% application in biophysics
% limited research in other system

The motion of a microscopic particle suspended in a medium provides information about  its interactions with its surroundings~\cite{Qian_1991,Saxton_1997}. It can be a purely random movement like the Brownian motion, but it can also be as complicated as hopping lipids on a cell membrane~\cite{Fujiwara_2002}. Measuring trajectories of single particles such as polystyrene beads~\cite{Beckerle_1984,Franosch_2011}, gold nanoparticles~\cite{Geerts_1987,Wu_2016f}, fluorescent proteins~\cite{Gahlmann_2014,Kusumi_2014}, and quantum dots~\cite{Chang_2008} have revealed fundamental processes and microscopic structures in nature~\cite{Ruthardt_2011,Manzo_2015,Spindler_2016,Taylor_2019a}. 
Thus, the observation of individual trajectories, known as single-particle tracking (SPT), has established itself as a powerful tool to study molecular processes and interactions, especially in biological systems~\cite{Saxton_1997,Shen_2017,Gahlmann_2014,Ruthardt_2011,Manzo_2015}. Moreover, SPT shows great promise for studying nanoscale chemical and physical processes~\cite{Franosch_2011,Araque_2015,Kirstein_2007,Higgins_2015}, and novel applications of SPT are still actively emerging~\cite{Shen_2017}.   

An optical tweezer is another useful tool in single-molecule studies, which provides a non-invasive way of manipulating microscopic objects with great precision~\cite{Grier_1997,Neuman_2004,Marago_2013,Bowman_2013,Jones_2015,Bustamante_2021}. It combines a Gaussian laser beam with a high numerical aperture objective lens to produce a tight focal spot, which creates an optical force that can bring small objects to the intensity maximum~\cite{Ashkin_1986}. Beyond the original capabilities to grab tiny objects~\cite{Ashkin_1987,Ashkin_1987a}, optical tweezers have found diverse applications enabled by its ability to locally apply precise and controlled forces, ranging from measuring piconewton forces in molecular motors~\cite{Svoboda_1993} to stretching or relaxing DNA to determine its elastic properties~\cite{Wang_1997}.

Optical tweezers can also facilitate the measurements of SPT by transporting or constraining a tracer particle to an area of interest. Although finding their initial and major applications in biophysics~\cite{Lang_2004,Comstock_2011,Bustamante_2021}, optical tweezers are also being increasingly implemented in physics such as microrheology~\cite{Yao_2009} and hydrodynamics~\cite{Polin_2006}. \citet{Franosch_2011}, for instance, studied the Brownian motion of an optically trapped bead in water and revealed that surrounding water molecules displaced by the particle's thermal motion act back on the particle, appearing as a non-white noise spectrum. In particular, the optical tweezer employed in 
their study made it possible to discover such weak interactions by providing controlled forces and thus facilitating the characterization of the particle motion.

Optical tweezer-assisted SPT, however, has not yet actively emerged in chemistry. The SPT itself has been extensively used in surface science to unveil molecular-level details of diffusion~\cite{Walder_2011,Skaug_2014}, mass transport~\cite{Higgins_2015}, catalytic reactions~\cite{Dong_2018}, and many other processes~\cite{Zhong_2020}, which were inaccessible with classical bulk or ensemble measurements~\cite{Atkins_2014a}. Optical trapping, on the other hand, has also found its own applications in surface science, such as in Raman spectroscopy~\cite{Yuan_2017b,Oyamada_2022} and photo-catalysis~\cite{Tsuboi_2021}. Despite these diverse implementations, the utilization of optical tweezers for SPT in colloid and interface science has not yet been accomplished. The abilities of optical tweezers to enable pinpoint access to the region of interest and SPT to monitor the time evolution of physical and chemical reactions at the molecular level hold great promises in surface science.  

Here, we demonstrate the application of optical tweezers in SPT analysis in colloid and interface science. We study a specific case of optically trapped gold nanoparticles, which also serve as an optical probe for SPT, interacting with surfactant molecules adsorbed at water--glass interfaces. Surfactants adsorbed on solid surfaces spontaneously form interesting structures such as spherical or cylindrical aggregates (also known as admicelles)~\cite{Ducker_1999,Tyrode_2008} and bilayer membranes~\cite{Evans_1987,Li_2020}. They exist both on the walls of liquid containers and the particles surfaces. We bring these surfaces close to each other by exclusively positioning a single particle near a glass wall using an optical tweezer and monitor the particle trajectories. Any changes in particle motions can then be related to the local interaction of the surfactant molecules near the particle. We show a qualitative correlation between the thermodynamic model of adsorption and the collective behaviors of the measured single-particle trajectories that are locally determined by their immediate vicinity. Furthermore, individual trajectories provide evidence of long-term interactions between adsorbed molecules on particle and glass surfaces achieved by stable and long trapping of the particles with the optical tweezer. Finally, we perform time-averaged statistical analyses on a single-particle trajectory, which further strengthen our interpretation of surfactant interactions.

\section{Materials and Methods}
\label{sec:methods}
\subsection{Materials}
Gold nanoparticles were used as an optical probe in this study. They scatter visible light resonantly, making their scattering cross sections much larger than their physical cross-sections and thus having been successfully utilized as nanoscale probes under optical microscopes~\cite{Geerts_1987,Kottmann_2001e,Wu_2016f,Spindler_2016}. They are also free from photobleaching or photoblinking, which enables long and continuous measurements unlike their fluorescence counterparts~\cite{Taylor_2019}. We purchased gold colloids with 150 nm diameter, stabilized in citrate buffer, from Sigma-Aldrich (a product of CytoDiagnostics, Inc.). 

On the other hand, cetyltrimethylammonium chloride (CTAC) was used as an example of cationic surfactants. We obtained CTAC from Sigma-Aldrich (25 wt \% in H$_2$O) and diluted the solution with distilled water to reach desired concentrations ranging from 0.1 mM to 2.0 mM. We prepared a separate gold nanoparticle-CTAC mixture for each CTAC concentration. The stock gold colloids were centrifuged for 30 min at 200 $g$ for separation and re-dispersed using a vortex mixer in a CTAC solution at a desired concentration. The concentration of gold nanoparticles was further reduced in order to have a sparse appearance under the optical microscope and thus ensure single-particle trapping.

Borosilicate glass coverslips from Menzel Gläser (145 $\mu m$ in thickness) and a double-sided adhesive spacer (120 $\mu m$  thickness, Grace Bio-Labs SecureSeal\textsuperscript{\texttrademark} imaging spacer) were used to build a fluid chamber. All the trapping experiments were performed inside the fluid chamber to keep the liquids from evaporating. 

\subsection{Preparation of glass surface}
We prepared two different types of glass surface to alter the adsorbed morphologies of the surfactant molecules. One type of glass is a native glass without any surface modification. The other type of glass is treated with oxygen plasma (at 200 W for 60 s with a 400 mL/min flowrate) to activate its surface~\cite{Alam_2014a}. The likely structures of adsorbed molecules and the effect of surface modification on them are discussed in the results and discussion section (Section~\ref{sec:surfactant morphology}). All glass coverslips were sonicated in acetone and isopropyl alcohol baths for 30 minutes each, before use or surface modification. 

\subsection{Optical trapping and video analysis of a trapped particle}
In each experiment, a single gold nanoparticle was optically trapped near a glass wall using a He-Ne laser. A 60$\times$, 0.85 NA, air objective lens was used to focus the trapping laser at the water--glass interface, and the laser power was controlled to be 10 mW before entering the trapping objective. The particle motion within an optical trap was recorded using a CMOS camera (CM3-U3-50S5C-CS, FLIR) at a 346 framerate. Its trajectory was traced using a python toolkit, Trackpy~\cite{Allan_2021}, which is based on a particle-tracking algorithm developed by~\citet{Crocker_1996}. More detailed information about the optical trapping setup and methods can be found in ref~\cite{Kim_2021e}. A dataset for all the particle recordings and the corresponding trajectories is available in our Zenodo data repository~\cite{Kim_2022}.

\section{Results and Discussion}
\label{sec:results}
\subsection{Surfactant adsorption at water--glass interface}~\label{sec:surfactant morphology}
Surfactant molecules adsorb at water--glass interfaces, spontaneously forming complex structures~\cite{Atkin_2003}. The self-assembly structures of the adsorbed molecules depend on several factors such as pH~\cite{Goloub_1996}, salt~\cite{Goloub_1997,Lamont_1998}, and surface preparation~\cite{Chorro_1999,Grant_2000}. The change of solution pH or salinity can alter the overall surfactant self-assemblies on both the particles surfaces and the glass walls. In contrast, the surface treatments on glass surfaces can provide a selective modification of adsorbed surfactants on the treated glass walls. Therefore, we focus on the latter to examine any changes in the particle--surface interactions controlled by surfactants conformations at the water--glass interface. 

\begin{figure}
    \centering
    \includegraphics[width=0.9\textwidth]{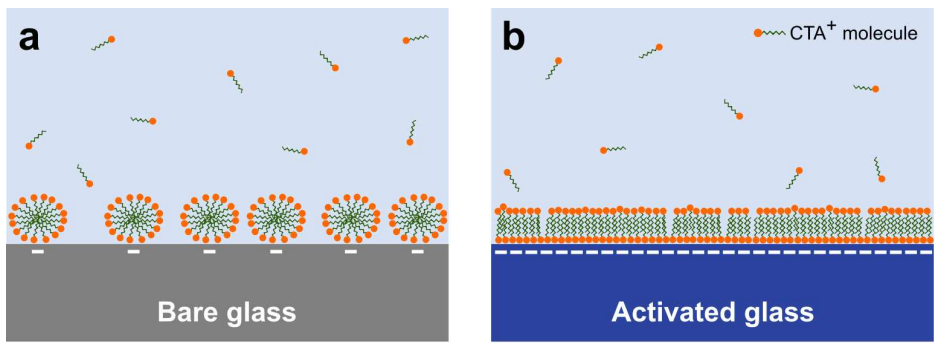}
    \caption{Schematic diagram showing probable morphologies of adsorbed surfactant molecules at the glass-water interface.
    \textbf{a}, A bare glass and
    \textbf{b}, a glass surface treated with oxygen (O$_2$) plasma. The activated glass surface has denser silanol groups compared to the native glass. These silanol (SiO$^-$) groups on each surface act as electrostatic binding sites for cationic surfactant adsorption.}
    \label{fig:schematic}
\end{figure}

We prepared two different glass surfaces to induce a morphological change in surfactant adsorption. The first type of glass surface is a native glass without any surface modification. In general, a glass surface becomes negatively charged when immersed in water through the deprotonation of its silanol groups~\cite{Behrens_2001}:
\begin{equation}
    \mathrm{SiOH} \rightleftharpoons \mathrm{SiO}^{-}+\mathrm{H}^{+} \,.
\end{equation}
The negatively charged silanol (SiO$^-$) groups drive the initial adsorption of CTAC through the electrostatic attraction with the cationic surfactant head groups (CTA$^+$). The adsorption of the molecules reduces the overall surface charge, and once the surface becomes neutralized, these molecules act as nucleation points for further adsorption~\cite{Atkin_2003}. At this stage, the hydrophobic interactions among surfactant tail groups drive the adsorption, similar to micelle formation in bulk solution. The molecules start to form centrosymmetric aggregates~\cite{Tyrode_2008}, aka admicelles, as shown in Figure~\ref{fig:schematic}a. The surface coverage rises steeply by forming these admicelles in this concentration span, and it soon reaches saturation near the critical micellar concentration (CMC).

The second type of glass surface is modified by oxygen-plasma treatment to increase the number of  silanol groups on its surface~\cite{Alam_2014a}. Henceforth, we call this type of glass an activated glass. After the plasma treatment, the activated surface becomes superhydrophilic due to the increased charged groups. Like the adsorption with native glasses, the electrostatic attraction drives the initial adsorption when the activated glass is immersed in surfactant solution. We postulate that the adsorbed molecules are so close together that they form a quasi-continuous film. In experiments, we observed an immediate increase in the contact angle with CTAC solution compared to the $0^\circ$ contact angle with water. Such increase in the contact angle implies that a surfactant monolayer is immediately formed with hydrophilic head groups electrostatically adsorbed to the surface and hydrophobic tail groups facing toward the bulk liquid. Any further adsorption that is hydrophobically driven forms a second layer above the first one (Figure~\ref{fig:schematic}b), rather than clustering as discrete aggregates as was the case for the bare glass.

Such changes in self-assembly morphologies induced by the surface properties have been extensively studied in the literature~\cite{Lamont_1998,Grant_2000,Ducker_1999}. \citet{Ducker_1999} studied surfactant (cetrimonium bromide, CTAB) aggregates on mica using AFM imaging. They found a transition from a flat bilayer to cylinders when reducing the binding sites on mica by introducing electrolyte (KBr). \citet{Lamont_1998} reported similar findings where they observed sequential changes in the CTAC structures on mica (bilayer $\rightarrow$ ordered cylinder $\rightarrow$ disordered cylinder $\rightarrow$ short cylinder $\rightarrow$ sphere) on the addition of rival cations (Cs$^+$). Lastly, \citet{Grant_2000} investigated the influence of surface hydrophobicity on the aggregate structures of nonionic surfactants where hydrophobic interactions are the main adsorption driving force. They observed the evolution of adsorbed structures from diffuse spherical aggregates to a monolayer with increasing surface hydrophobicity. All these findings deliver the same message that the increase in active binding sites on surfaces leads to lower-curvature structures. Based on these ideas, we postulate that CTAC forms a bilayer structure on activated glass and discrete globular aggregates on bare glass as illustrated in Figure~\ref{fig:schematic}.

\subsection{Particle trajectory in a harmonic optical trap}
\begin{figure}[p!]
    \centering
    \includegraphics[width=\textwidth]{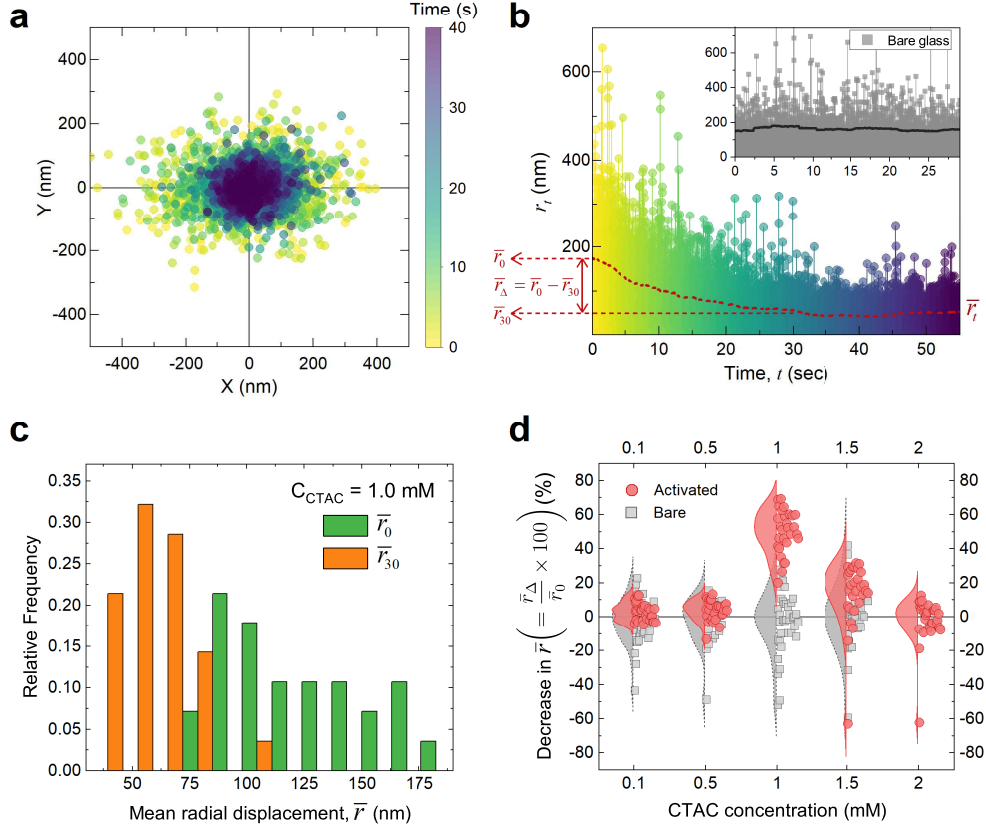}
    \caption{Time-varying motion of particles in harmonic optical traps.
    \textbf{a}, Position distribution of a single optically-trapped particle near an activated glass wall, color-mapped with time.
    \textbf{b}, Radial displacements (defined in Equation~\ref{eqn:r}) as a function of time for the same particle in \textbf{a}. The red-dashed line indicates a moving average (${{\bar r}_t}$) with a subset size of 1730 data points. The inset shows the radial displacement of the same kind of particle at the water/bare glass interface for comparison; the grey solid line shows the same moving average. 
    \textbf{c}, Histograms of the mean radial displacements at $t = 0$ and 30 s. From \textbf{a} to \textbf{c}, the CTAC concentration was 1.0 mM, and the glass surface was activated by O$_2$-plasma treatment.
    \textbf{d}, Comparison of the decreases in $\bar r$ for the first 30 seconds between the bare and activated surfaces as a function of CTAC concentration. The reduced amount, ${{\bar r}_\Delta}$, is expressed as a percentage of its initial value, ${{\bar r}_0}$.
    }
    \label{fig:decrease}
\end{figure}

An optical tweezer was employed in this study to confine the motion of a particle in a local domain and facilitate monitoring the particle's trajectory. Figure~\ref{fig:decrease}a shows an example time-series position data extracted from a video recording. (An example image and feature-finding process are illustrated in Supplementary Figure~\ref{SI:locating}. More example trajectories are shown in Supplementary Figure~\ref{SI:pattern}.) Each position data point is color-mapped, gradually changing from yellow to dark blue for legibility. The position fluctuation inside the optical trap decreases over a timespan of 40~s in Figure~\ref{fig:decrease}a. Such changes with time deviates from a simple diffusion and can be related to the interaction that the particle undergoes with its surroundings, especially with the molecules adsorbed on the glass wall.

To quantitatively assess the time-dependent position data, we defined a radial displacement, $r_t$, as the distance from the origin (corresponding to the trap center) to the particle position $(X(t),Y(t))$:
\begin{equation}~\label{eqn:r}
    r_t = \sqrt {{X(t)^2} + {Y(t)^2}} \,.
\end{equation}
Figure~\ref{fig:decrease}b shows the evolution of $r_t$ for the same particle trajectory in Figure~\ref{fig:decrease}a. The gradual decrease in the envelope of $r_t$  depicts well the temporal changes in the particle motion inside the optical trap. Since $r_t$ fluctuates between its extremes, we calculated the moving average of $r_t$ to smooth out the short-term fluctuations and quantify its long-term decrease. We used a time window of 5 s (corresponding to 1730 data points) to calculate the moving average, ${{\bar r}_t}$, and plotted it as the red-dashed line in Figure~\ref{fig:decrease}b. 

Interestingly, such changes in particle motions occur only with activated glass surfaces. For a bare glass substrate at the same CTAC concentration, shown in the inset of Figure~\ref{fig:decrease}b, there is no apparent change in the average radial displacement, ${{\bar r}_t}$ (the solid grey line in the inset). This difference between the native and activated glasses indicates that the structure of the adsorbed surfactants at the water--glass interface plays a key role in the particle--surface interactions.

This hypothesis that the particle--surface interaction is sensitive to the surfactant morphologies can be tested in two ways: (1) comparing the behavior of ${{\bar r}_t}$ between bare and activated glass surfaces and (2) varying the CTAC concentration which provides different adsorption states, \textit{e.g.}, surface coverage and aggregate structures. To facilitate the comparisons between the different surfaces and among different concentrations, we quantified the amount of  long-term decrease by the difference in the value of ${{\bar r}_t}$ during the first 30 s of trapping:
\begin{equation}
    {{\bar r}_\Delta} = {{\bar r}_0} - {{\bar r}_{30}} \,,
\end{equation}
which is also indicated as a red arrow in Figure~\ref{fig:decrease}b. We chose the time window of 30 s as it is a typical length of each recording, and (not all but) most of the particles reach a steady state within this period.

We performed dozens of measurements for each of the experimental conditions to average out the effect of experimental variations such as variances in particle sizes and/or charges, and the inhomogeneity of glass surfaces. Different trajectories in Supplementary Figure~\ref{SI:pattern} demonstrate these variances within the same experimental condition. Figure~\ref{fig:decrease}c illustrates an example analysis for 28 measurements at 1.0 mM CTAC with activated surfaces. The ${{\bar r}_0}$ represents the average displacement immediately after trapping, whereas the ${{\bar r}_{30}}$ represents the average displacement after staying 30~s in the optical trap. The overall left-shift of the histogram toward smaller values indicates that the repetitive measurements at different surface locations with randomly picked particles converges to similar confined trajectories. In other words, all the particles experienced more restrictions in their motions with time evolution. 

We tested five different concentrations (0.1, 0.5, 1.0, 1.5, and 2.0 mM), which cover the range below and above the critical micellar concentration (CMC) so that each concentration corresponds to different adsorption phases. In general, the extent of surface coverage reaches its maximum near the CMC~\cite{Atkin_2003, Tyrode_2008}. We found that the CMC of our particle-CTAC mixture is 1.4 mM (Supplementary Figure~\ref{SI:zeta}a), which is higher than the values reported in the literature (1.0 to 1.1 mM) due to the existence of the gold colloids in the solution. We refer to our recent publication for more details~\cite{Kim_2021e}. 

Figure~\ref{fig:decrease}d summarizes the analyses of the amount of the decrease in ${{\bar r}}$ for bare and activated substrates as a function of CTAC concentration. The reduction in the mean radial displacement, ${{\bar r}_\Delta}$, is expressed as a percentage of its initial value, ${{\bar r}_0}$. For each concentration, the data points ($32 \pm 6$ measurements on average) are shown on the right, and the distribution fit is shown on the left. For all the examined concentrations, the average temporal change is absent for the bare glass cases, with a distribution centered around the origin as also seen in the inset in Figure~\ref{fig:decrease}b. It implies that regardless of the surface coverage increase with the concentration, the spherical admicelles on bare glass substrates have no influence on the long-term changes in particle motion. 

On the other hand, the average decay in ${{\bar r}}$ is much noticeable for the activated glasses but only at specific concentrations, especially 1 mM and 1.5 mM in Figure~\ref{fig:decrease}d. These concentration-dependent interactions bolster the proposed hypothesis that the particle--surface interaction is responsive to the surfactant morphologies. We will elaborate on the effect of surface coverage and adsorbed morphologies in the following section.

\subsection{Effects of surfactant conformations on the particle--surface interactions}~\label{sec:evolution}
The stark contrast of the particle motion between bare and activated surfaces proves that the arrangement of the surfactant molecules at the water--glass interface is crucial to determine whether or not long-term particle--surface interactions occur. One remaining question is why these interactions occur only with the bilayer structures and not with the globular aggregates. 

To answer this question, we also need to consider the surface of the particles. The particle surface before mixing with the surfactant solution is initially covered with citrate anions, used as a stabilizing agent. On the addition of surfactant solution, the CTA$^+$ cations adsorb on the citrate-capped gold surfaces. According to a recent study by \citet{Li_2020}, the self-assembly structure of CTA$^+$ cations on citrate-capped gold surfaces undergoes, with increasing concentration, a gradual transition from a monolayer to a bilayer. They also found that the particles readily aggregated if the surfactant self-assembly formed monolayers, due to the hydrophobic interactions among the hydrocarbon tails exposed to the bulk fluid. Since we did not find any clue for aggregation in our size measurement using dynamic light scattering, we postulated that the surfactant molecules form bilayer structures on the particle surfaces across the concentration range we studied. 

The first layer of this bilayer consists of surfactant molecules adsorbed to the gold surface by electrostatic attraction. The second layer comprises molecules adsorbed to the first layer by hydrophobic interactions (similar to the bilayer formation on a flat surface). For this bilayer formation, the completion of the first layer precedes the initiation of the second layer~\cite{Li_2020}. The fractional coverage of the second layer then gradually increases with concentration. We experimentally confirmed the gradual growth of the second layer by measuring the concomitant increase in the particles' zeta potential in Supplementary Figure~\ref{SI:zeta}b. Once the coverage of the second layer reaches its maximum and the surfactant concentration is higher than the CMC, micelles in the bulk fluid can also be associated with the particle surface~\cite{Li_2020}. By measuring the hydrodynamic size of the particles, we discovered the involvement of adsorbed micelles near the particles. In the Supplementary Figure ~\ref{SI:zeta}c, the abrupt decrease in the hydrodynamic size above the CMC indicates a decrease in the thickness of the electric double layer due to the increase in local ionic concentration. Therefore, it can be interpreted as the presence of micelles strongly associated with the particle surface.

To summarize, the surfactant bilayers present at both the water--glass interface and the particle surface are key to understanding their interactions. Their structural resemblance, for instance, can make them more likely to merge, similar to the fusion of lipid bilayers in life~\cite{Helm_1992,Petrache_1998}. In general, bilayers are known as a highly mobile structure~\cite{Atkins_2014}. In the case of the bilayers in our study, the first layers electrostatically adsorbed onto the surface are tightly bound and not easily dissociated with the surface. In contrast, the molecules comprising the second layers are ceaselessly twisting and writhing and can migrate over the first layer without leaving the surface completely~\cite{Atkin_2003}. Therefore, when the two bilayers come close in the instances of particle trapping, a disruption and rearrangement of the outer layers may occur, which can be observed as changes in particle behaviors.

\begin{figure}
    \centering
    \includegraphics[width=\textwidth]{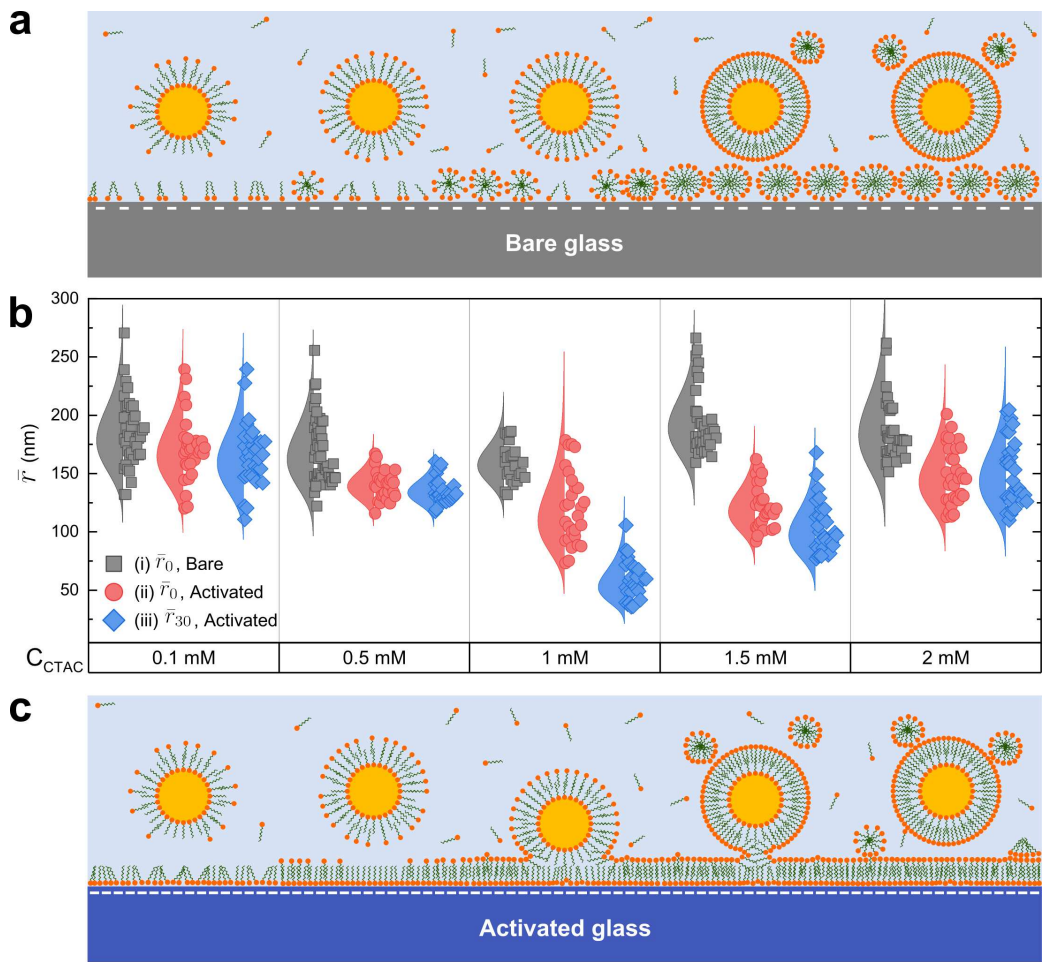}
    \caption{Effects of surfactant morphologies on particle motion as a function of concentration. 
    \textbf{a} and \textbf{c}, Cartoons showing plausible surfactant self-assembly formations at the gold/water and glass/water interfaces. When the glass surface is activated with O$_2$-plasma treatment, the adsorbed surfactant structure is assumed to undergo a conformational change from globular aggregates to bilayers. The changes in surface coverage and aggregate morphologies is depicted according to the surfactant concentration.   
    \textbf{b}, Distributions of mean radial displacement (${{\bar r}}$) as a function of CTAC concentration. Three datasets are displayed side by side for comparison: (i) $\bar r_0$, bare glass, (ii) $\bar r_0$, activated glass, and (iii) $\bar r_{30}$, activated glass.
    }
    \label{fig:comparison}
\end{figure}

Figure~\ref{fig:comparison}a and c show schematic representations of the surfactant structures at the water--glass interface and at the gold nanoparticle surface for the two different substrates. They illustrate the likely development of the surfactant conformations over the concentration range where the particle trajectories were investigated. Figure~\ref{fig:comparison}b shows the corresponding trajectory analyses, quantified as the statistics of the mean radial displacement, ${{\bar r}_t}$, as a function of CTAC concentration. Three datasets are plotted side by side for comparison: (i) ${{\bar r}_0}$ for bare glass, (ii) ${{\bar r}_0}$ for activated glass, and (iii) ${{\bar r}_{30}}$ for activated glass.  The distribution of ${{\bar r}_{30}}$ for bare glass is not shown in Figure~\ref{fig:comparison}b for clarity as it remains similar to (i), \textit{i.e.}, no time-dependence in ${{\bar r}_t}$ for bare glass as indicated in Figure~\ref{fig:decrease}d. We also refer the readers to our previous work~\cite{Kim_2021e} for more details about the particle interactions with spherical admicelles on bare glass surfaces. Here we focus on the difference between bare and activated surfaces, and especially, the particles' behaviors on bilayer membranes.

A comparison among (i), (ii), and (iii) across different concentrations provides qualitative insights into the mechanisms that govern the particles motions: one is the hydrophobic interaction between the particle and the surface, and the other is the extent of bilayer coverage. Compared to admicelles, the primary distinction of bilayer structure is that the hydrophobic interiors can be exposed to the bulk fluid. The exposed hydrophobic groups on a glass surface can interact with hydrophobic groups on a particle surface when a particle approaches. The resulting attraction force explains the decrease in $\bar r$ for the bilayer structures in Figure~\ref{fig:comparison}b. An exception occurs, however, at the lowest concentration (0.1 mM), where the structural difference at this low coverage is negligible.
 
On the other hand, the coverage extent determines how much and how fast the $\bar r$ decrease will be. When a particle approaches, the hydrophobic interaction brings the particle closer to the surface. As a result, the charged head groups can be disrupted and rearrange each other, creating a hydrophobic patch below the particle (as illustrated below the second particle in Figure~\ref{fig:comparison}c) and, consequently, developing an electrostatic trapping potential. The size of the hydrophobic patch will determine how tight the confinement will be. At low coverages, the confinement of the particle will be moderate, and the reaction will take place promptly as there are a  relatively small number of molecules to rearrange. The results at 0.5 mM in Figure~\ref{fig:comparison}b support this idea. They show a relatively modest decrease in $\bar r$ compared to 1 mM or 1.5 mM; furthermore,  this decrease happens in the early stage of the trapping so that we do not see much difference between $\bar r_0$ and $\bar r_{30}$. The variance of the distribution of $\bar r$ also decreases significantly, compared to those at the lowest concentration. In other words, the rearrangement process seems to reduce the effect of experimental variations coming from surface heterogeneity.

As the coverage increases, the overall hydrophobic area shrinks, and the surfactant bilayer surrounds more tightly a trapped particle, as depicted in Figure~\ref{fig:comparison}c. The dramatic decrease in $\bar r$ at 1.0 mM in Figure~\ref{fig:comparison}b supports this explanation. Another noticeable difference at this concentration is the steady decrease in $\bar r$ over tens of seconds (from $\bar r_0$ to $\bar r_{30}$) and the broadened distribution of $\bar r_0$. As shown in Supplementary Figure~\ref{SI:pattern}, the time it takes to reach an equilibrium varies with each measurement due to the random nature of adsorption. As a result of the varying rates, the distribution of $\bar r$, especially the initial averages ($\bar r_0$), is significantly broadened. The most likely explanation for the slow interaction, on the other hand, is that the incomplete particle bilayer is first combined with the nearest hydrophobic patches on the surface (the initial response related to $\bar r_0$), and this contact domain gradually grows by taking in other hydrophobic patches that were initially not in the immediate vicinity but slowly diffused over the surface to the particle location (the transition from $\bar r_0$ to $\bar r_{30}$)~\cite{Walder_2011}. Such a slow diffusion is less obvious at 1.5 mM in Figure~\ref{fig:comparison}b. This can be interpreted that the bilayer coverage has almost reached its saturation at this concentration near the CMC (1.4 mM), and there are hardly any exposed hydrophobic interiors nearby that can diffuse and cooperate.

At the highest concentration (2.0 mM) above the CMC, the effect of bilayer interaction is further reduced. The particles' zeta potentials have increased due to micelles' association with the particle surface (Supplementary Figure~\ref{SI:zeta}). The stronger electrostatic repulsion between the particle and the surface can consequently hamper the interaction (the last particle in Figure~\ref{fig:comparison}c). Accordingly, the $\bar r$ recovers its mean and variance comparable to those at the lowest concentration.

With the optical tweezer, which provides a facile manipulation of particle location with the assistance of precise piezoelectric control, we also observed a particle motion while we disturbed the stable trapping by relocating the particle to a new place and bringing it back a few seconds later. Supplementary video 1 shows such a particle recording at 1 mM CTAC, where the long-term particle--surface interaction is most apparent. At the beginning of the recording, it shows a particle stably trapped after reaching its steady state (\textit{i.e.}, an equilibrium) at the center. At around 3~s of the timestamp in the video, the particle is moved to a new location on the left side where it becomes immediately destabilized and settles down slowly, similar to our previous observations at this CTAC concentration. The particle stays in this new trap position for about 8~s and is moved back to the original location. Surprisingly, at the original trap location, the particle instantly recovers its stabilized motion. Such an immediate recovery attests to the depletion of the surfactant bilayer at the optical trap, which remains even after the particle leaves the place. It also implies that the back diffusion of molecules to this depleted area is reciprocally a slow process.

\subsection{Statistical analyses of a time-varying single-particle trajectory}

We have interpreted particle dynamics by examining particle trajectories and investigating the role of surfactant morphologies in particle--surface interactions. So far, we have focused on the ensemble behaviors and their variances changing with time and concentration. Compared to these statistical ensembles, each trajectory can also be analyzed further by taking the statistics along its time axis. The \textit{time-averaged} mean squared displacement (MSD) is the most common measure when we deal with particle trajectories. Using the time-averaged MSD, for example, we can analyze the forces acting on the particle, based on \textit{a priori} knowledge of particle dynamics~\cite{Qian_1991,Norrelykke_2011}. To be more precise, we first formulate a Langevin equation to describe the motion of a particle and compare the measured MSD with the MSD solved for the Langevin equation. Any deviation from the theoretically solved MSD can prove the involvement of additional forces or interactions.

We modeled a particle in the optical trap as a damped harmonic oscillator in a fluid with a Langevin equation~\cite{Norrelykke_2011}: 
\begin{equation}
    m\ddot x(t) + \gamma \dot x(t) + \kappa x(t) = {F_{therm}}(t)\,, \label{eqn:Langevin}
\end{equation}
where $m$ is its inertial mass, $\gamma$ is its damping (fluid friction) coefficient, $\kappa$ is the spring constant of the optical trap, and $F_{therm}$ is the random thermal force. The inertial term, $m\ddot x(t)$, can be omitted from the equation for describing a particle in an aqueous medium on the timescale longer than a few microseconds~\cite{Purcell_1977}. The solution for MSD to the Langevin equation without the inertial term becomes~\cite{Norrelykke_2011}:
\begin{equation}
    {\mathop{\rm MSD_{Langevin}}\nolimits} (\tau ) = \frac{{2{k_B}T}}{\kappa }\left( {1 - {e^{ - \tau \kappa /\gamma }}} \right) \,, \label{eqn:MSD_sol}
\end{equation}
where $k_B$ is the Boltzmann constant and $T$ the temperature.  

To attest to the involvement of additional forces arising from the particle--surface interactions, we calculated the time-averaged MSD from a measured trajectory shown in Figures~\ref{fig:decrease}a and b.  
This trajectory, ${{\bf{T}}_N}$, is a time-series position data of length $N$:
\begin{equation}
    {{\bf{T}}_N} = \{ ({X_1},{Y_1}),({X_2},{Y_2}), \ldots ,({X_N},{Y_N})\}\,,
\end{equation}
where each position is defined in a two-dimensional space. For this trajectory, ${{\bf{T}}_N}$, the time-averaged MSD is defined as follows~\cite{Qian_1991}:
\begin{equation}~\label{eqn:ta-msd}
    {\rm{MSD_{measured}}}(\tau ) = \frac{1}{{N - \tau }}\sum\limits_{i = 1}^{N - \tau } {\{ {{({X_{i + \tau }} - {X_i})}^2} + {{({Y_{i + \tau }} - {Y_i})}^2}} \}\,,
\end{equation}
for any time lag $\tau = 1,2,...,N-1$. We subdivided the trajectory into five second-long segments and calculated the time-averaged MSD for the first segment ($t = 0 - 5$ s) and another segment after reaching the steady-state ($t = 50 - 55$ s) for comparison. 

\begin{figure}
    \centering
    \includegraphics[width=\textwidth]{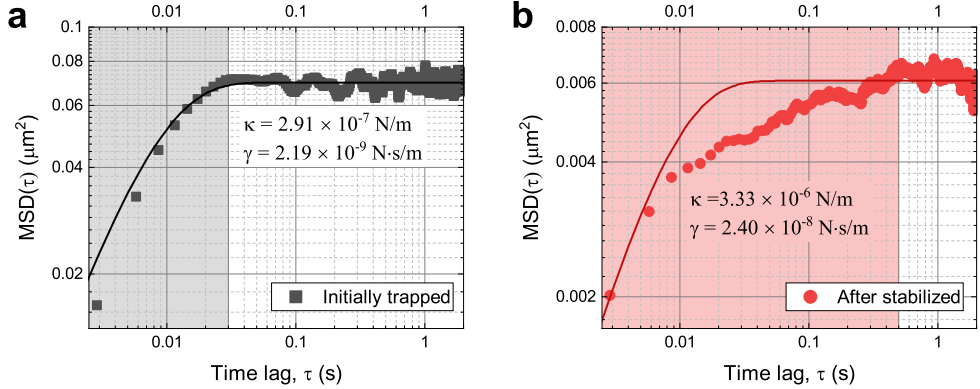}
    \caption{Time-averaged mean squared displacement (MSD) of segmented trajectories. \textbf{a}, Time-averaged MSD calculated from the first five-second segment of the trajectory shown in Figure~\ref{fig:decrease}a. The MSD solution of the Langevin equation in 2D (black solid line) is fitted to the measured MSD (grey squares). 
    \textbf{b} Time-averaged MSD from a segment of the same trajectory after reaching a steady state (red circles). The 2D solution of the Langevin equation (red solid line) is fitted to the data, with constraints on $\kappa$ and $\gamma$ to keep the fitted curve within the upper and lower limits of the measured MSD. 
    The shaded areas in \textbf{a} and \textbf{b} indicate the time lags to reach the MSD plateaus.
    }
    \label{fig:msd}
\end{figure}

The first segment corresponds to the timespan when the particle has just been trapped and initiated any interactions. The MSD calculated from this segment using Equation~\ref{eqn:ta-msd} is plotted in Figure~\ref{fig:msd}a in a double logarithmic scale. The next step is to fit the MSD\textsubscript{Langevin} to this MSD obtained from the experiment. However, the solution given in Equation~\ref{eqn:MSD_sol} is for a one-dimensional trajectory. Since we used the measured trajectory defined in two dimension (Equation~\ref{eqn:ta-msd}), the solution of the Langevin equation should also be multiplied by the  trajectory dimension. Therefore, we used a modified solution of the Langevin equation for two dimension ($=2\times{\mathop{\rm MSD_{Langevin}}\nolimits} (\tau )$) to fit it to the measured MSD. Before executing the curve fitting, we also estimated the temperature $K$ at the particle to be 366.61 K, a value calculated from an electromagnetic heating simulation (Supplementary Section~\ref{SI:heating} for more details). Then we estimated the values of the spring ($\kappa$) and damping ($\gamma$) constants in the least-squares sense as shown in Figure~\ref{fig:msd}a. The measured MSD shows a good agreement with the theoretical MSD, implying that the Langevin equation successfully describes the initially trapped particle motions. 

Since we know the average size of the particle, we can also roughly estimate the drag coefficient $\gamma$ using Stokes' drag (the frictional force exerted on a spherical object in a viscous fluid): $\gamma_{Stokes} = 3\pi \eta d$, where $\eta$ is the viscosity of the fluid and $d$ the particle diameter. This drag should be corrected further as the particle moves near the glass wall~\cite{Happel_2012}. Assuming that the particle--surface gap is roughly 1/10 of the particle radius, the hydrodynamic drag increases by $\sim$2.36~times near the surface compared to the Stokes' drag~\cite{Svoboda_1994a}. The drag coefficient, therefore, is approximated as $1.41\times10^{-9}$~N$\cdot$s/m by assuming the particle diameter $d=150$ nm and the viscosity $\eta = 0.423$ mPa$\cdot$s for water viscosity at 340 K (Supplementary Section~\ref{SI:heating} and Figure~\ref{SI:temp} for the water temperature simulation). This value of $\gamma$ is comparable to the value we found from the fitting of MSD$_{\rm{Langevin}}$ to the measured MSD ($2.19\times10^{-9}$~N$\cdot$s/m). The discrepancy between them can be attributed to any mismatch between the assumption and the actual particle size and water viscosity values. It can also be attributed to the particle--surface interaction during the first five trapping seconds. 

The MSD calculated from the segment after reaching the steady-state, however, exhibits anomalous behavior that is significantly deviating from theoretical predictions as shown in Figure~\ref{fig:msd}b. For the curve fitting in this case, we applied a boundary condition for $\kappa$ and $\gamma$ ($0 < \kappa  < 4 \times {10^{ - 6}}$; $0 < \gamma  < 2.4 \times {10^{ - 8}}$) to keep the fitted curve between the minimum and maximum values of the measured MSD for the measured time lags. The estimated value for $\kappa$ has increased about ten times compared to the value fitted to the initial segment. It supports our hypothesis that the rearrangement of the incomplete bilayer can create complementary trapping potentials. The estimated drag, $\gamma$, has also increased about ten times, likely due to the increased restrictions in particle motion imposed by the bilayer interactions based on our hypothesis. Most importantly, the Langevin equation now fails to describe the motion of the particle after reaching the steady-state. This implies that the particle does not behave like a damped harmonic oscillator anymore, suggesting the emergence of a new type of force or physical interaction such as the fusion of the bilayers, as we suggested earlier. 

Based on theoretical works~\cite{Neusius_2009,Jeon_2010}, the subdiffusive behavior between $\tau=0.01$ s and 0.5 s in Figure~\ref{fig:msd}b can represent geometrical confinement, meaning that the particle is restricted to a finite domain. For example, several experimental findings in biology also suggested that the restricted motion of a probe molecule inside a cell can arise due to polymer networks in the cytoplasm~\cite{Toli_2004} or the presence of the cell nuclei~\cite{Wachsmuth_2000}. Furthermore, a very similar time-averaged MSD dependence on $\tau$ was also reported by \citet{Jeon_2013} with a polystyrene bead in a worm-like micellar solution, where the elongated micelles form a transient polymer network. They explained that such a characteristic subdiffusive motion in the millisecond range resembles the movement of a particle in a viscoelastic medium. All these theoretical and experimental findings provide solid evidence that the particle in Figure~\ref{fig:msd}b experiences geometrical confinement and a coincident viscoelastic effect, which can be straightforwardly related to the interactions among surfactant molecules. However, in this case, these interactions must have been constrained to the interface since the surfactant concentration was below the CMC, \textit{i.e.}, prior to the micelle formation in the bulk solution. Therefore, the time-averaged MSD corroborates the rearrangement and partial fusion of the layers proposed in the previous section.

\section{Conclusions}
We have examined the trajectories of colloidal particles locally confined at a water--glass interface by an optical tweezer. These optically bright particles acted as nanoscale probes that could respond to conformational changes in surfactant self-assemblies adsorbed at the solid--liquid interfaces. These conformational changes include the development of surface coverages during the adsorption process, morphological changes induced by surface preparation, and dynamic interactions between particles and surfaces. The ensemble statistics of single-particle trajectories provided information on the average behavior of the particles, which was well correlated with the adsorption phase described in the literature~\cite{Atkin_2003}. Individual trajectories, on the other hand, provided information about the dynamic arrangement of the adsorbed molecules in real time, which have been obscured in traditional thermodynamic experiments. We combined these spatial and temporal evidences and provided new insights into the interactions governing the particle motion. In particular, by studying the special system consisting of a gold nanoparticle, a glass--water interface, and surfactants, we found evidence of the fusion of surfactant bilayers in the narrow gap between the particle and the surface. This interpretation was further supported by statistical analysis on the individual particle trajectory evaluated on the time axis. Our findings extend our fundamental understanding of surfactant dynamics on colloid and solid surfaces and demonstrate the great potential of optically trapped nanoparticles as probes to investigate chemical processes at the nanoscale.

\section{Acknowledgements}
Funding from the European Research Council (ERC-2015-AdG-695206 Nanofactory) is gratefully acknowledged.

%% The Appendices part is started with the command \appendix;
%% appendix sections are then done as normal sections

%% If you have bibdatabase file and want bibtex to generate the
%% bibitems, please use
%%
\bibliographystyle{elsarticle-num-names} 
\biboptions{sort&compress}
\bibliography{references}

\begin{thebibliography}{71}
\expandafter\ifx\csname natexlab\endcsname\relax\def\natexlab#1{#1}\fi
\providecommand{\url}[1]{\texttt{#1}}
\providecommand{\href}[2]{#2}
\providecommand{\path}[1]{#1}
\providecommand{\DOIprefix}{doi:}
\providecommand{\ArXivprefix}{arXiv:}
\providecommand{\URLprefix}{URL: }
\providecommand{\Pubmedprefix}{pmid:}
\providecommand{\doi}[1]{\href{http://dx.doi.org/#1}{\path{#1}}}
\providecommand{\Pubmed}[1]{\href{pmid:#1}{\path{#1}}}
\providecommand{\bibinfo}[2]{#2}
\ifx\xfnm\relax \def\xfnm[#1]{\unskip,\space#1}\fi
%Type = Article
\bibitem[{Qian et~al.(1991)Qian, Sheetz, and Elson}]{Qian_1991}
\bibinfo{author}{H.~Qian}, \bibinfo{author}{M.~Sheetz},
  \bibinfo{author}{E.~Elson},
\newblock \bibinfo{title}{Single particle tracking. analysis of diffusion and
  flow in two-dimensional systems},
\newblock \bibinfo{journal}{Biophys. J.} \bibinfo{volume}{60}
  (\bibinfo{year}{1991}) \bibinfo{pages}{910--921}.
%Type = Article
\bibitem[{Saxton and Jacobson(1997)}]{Saxton_1997}
\bibinfo{author}{M.~J. Saxton}, \bibinfo{author}{K.~Jacobson},
\newblock \bibinfo{title}{Single-particle tracking: Applications to membrane
  dynamics},
\newblock \bibinfo{journal}{Annu. Rev. Biophys. Biomol. Struct.}
  \bibinfo{volume}{26} (\bibinfo{year}{1997}) \bibinfo{pages}{373--399}.
%Type = Article
\bibitem[{Fujiwara et~al.(2002)Fujiwara, Ritchie, Murakoshi, Jacobson, and
  Kusumi}]{Fujiwara_2002}
\bibinfo{author}{T.~Fujiwara}, \bibinfo{author}{K.~Ritchie},
  \bibinfo{author}{H.~Murakoshi}, \bibinfo{author}{K.~Jacobson},
  \bibinfo{author}{A.~Kusumi},
\newblock \bibinfo{title}{Phospholipids undergo hop diffusion in
  compartmentalized cell membrane},
\newblock \bibinfo{journal}{J. Cell Biol.} \bibinfo{volume}{157}
  (\bibinfo{year}{2002}) \bibinfo{pages}{1071--1082}.
%Type = Article
\bibitem[{Beckerle(1984)}]{Beckerle_1984}
\bibinfo{author}{M.~C. Beckerle},
\newblock \bibinfo{title}{Microinjected fluorescent polystyrene beads exhibit
  saltatory motion in tissue culture cells},
\newblock \bibinfo{journal}{J. Cell Biol.} \bibinfo{volume}{98}
  (\bibinfo{year}{1984}) \bibinfo{pages}{2126}.
%Type = Article
\bibitem[{Franosch et~al.(2011)Franosch, Grimm, Belushkin, Mor, Foffi, Forró,
  and Jeney}]{Franosch_2011}
\bibinfo{author}{T.~Franosch}, \bibinfo{author}{M.~Grimm},
  \bibinfo{author}{M.~Belushkin}, \bibinfo{author}{F.~M. Mor},
  \bibinfo{author}{G.~Foffi}, \bibinfo{author}{L.~Forró},
  \bibinfo{author}{S.~Jeney},
\newblock \bibinfo{title}{Resonances arising from hydrodynamic memory in
  brownian motion},
\newblock \bibinfo{journal}{Nature} \bibinfo{volume}{478}
  (\bibinfo{year}{2011}) \bibinfo{pages}{85--88}.
%Type = Article
\bibitem[{Geerts et~al.(1987)Geerts, De~Brabander, Nuydens, Geuens, Moeremans,
  De~Mey, and Hollenbeck}]{Geerts_1987}
\bibinfo{author}{H.~Geerts}, \bibinfo{author}{M.~De~Brabander},
  \bibinfo{author}{R.~Nuydens}, \bibinfo{author}{S.~Geuens},
  \bibinfo{author}{M.~Moeremans}, \bibinfo{author}{J.~De~Mey},
  \bibinfo{author}{P.~Hollenbeck},
\newblock \bibinfo{title}{Nanovid tracking: A new automatic method for the
  study of mobility in living cells based on colloidal gold and video
  microscopy},
\newblock \bibinfo{journal}{Biophys. J.} \bibinfo{volume}{52}
  (\bibinfo{year}{1987}) \bibinfo{pages}{775}.
%Type = Article
\bibitem[{Wu et~al.(2016)Wu, Lin, Yen, and Hsieh}]{Wu_2016f}
\bibinfo{author}{H.-M. Wu}, \bibinfo{author}{Y.-H. Lin}, \bibinfo{author}{T.-C.
  Yen}, \bibinfo{author}{C.-L. Hsieh},
\newblock \bibinfo{title}{Nanoscopic substructures of raft-mimetic
  liquid-ordered membrane domains revealed by high-speed single-particle
  tracking},
\newblock \bibinfo{journal}{Sci. Rep.} \bibinfo{volume}{6}
  (\bibinfo{year}{2016}) \bibinfo{pages}{20542}.
%Type = Article
\bibitem[{Gahlmann and Moerner(2014)}]{Gahlmann_2014}
\bibinfo{author}{A.~Gahlmann}, \bibinfo{author}{W.~E. Moerner},
\newblock \bibinfo{title}{Exploring bacterial cell biology with single-molecule
  tracking and super-resolution imaging},
\newblock \bibinfo{journal}{Nat. Rev. Microbiol.} \bibinfo{volume}{12}
  (\bibinfo{year}{2014}) \bibinfo{pages}{9}.
%Type = Article
\bibitem[{Kusumi et~al.(2014)Kusumi, Tsunoyama, Hirosawa, Kasai, and
  Fujiwara}]{Kusumi_2014}
\bibinfo{author}{A.~Kusumi}, \bibinfo{author}{T.~A. Tsunoyama},
  \bibinfo{author}{K.~M. Hirosawa}, \bibinfo{author}{R.~S. Kasai},
  \bibinfo{author}{T.~K. Fujiwara},
\newblock \bibinfo{title}{Tracking single molecules at work in living cells},
\newblock \bibinfo{journal}{Nat. Chem. Biol.} \bibinfo{volume}{10}
  (\bibinfo{year}{2014}) \bibinfo{pages}{524}.
%Type = Article
\bibitem[{Chang et~al.(2008)Chang, Pinaud, Antelman, and Weiss}]{Chang_2008}
\bibinfo{author}{Y.-P. Chang}, \bibinfo{author}{F.~Pinaud},
  \bibinfo{author}{J.~Antelman}, \bibinfo{author}{S.~Weiss},
\newblock \bibinfo{title}{Tracking bio-molecules in live cells using quantum
  dots},
\newblock \bibinfo{journal}{J. Biophotonics} \bibinfo{volume}{1}
  (\bibinfo{year}{2008}) \bibinfo{pages}{287}.
%Type = Article
\bibitem[{Ruthardt et~al.(2011)Ruthardt, Lamb, and Bräuchle}]{Ruthardt_2011}
\bibinfo{author}{N.~Ruthardt}, \bibinfo{author}{D.~C. Lamb},
  \bibinfo{author}{C.~Bräuchle},
\newblock \bibinfo{title}{Single-particle tracking as a quantitative
  microscopy-based approach to unravel cell entry mechanisms of viruses and
  pharmaceutical nanoparticles},
\newblock \bibinfo{journal}{Mol. Ther.} \bibinfo{volume}{19}
  (\bibinfo{year}{2011}) \bibinfo{pages}{1199--1211}.
%Type = Article
\bibitem[{Manzo and Garcia-Parajo(2015)}]{Manzo_2015}
\bibinfo{author}{C.~Manzo}, \bibinfo{author}{M.~F. Garcia-Parajo},
\newblock \bibinfo{title}{A review of progress in single particle tracking:
  from methods to biophysical insights},
\newblock \bibinfo{journal}{Rep. Prog. Phys.} \bibinfo{volume}{78}
  (\bibinfo{year}{2015}) \bibinfo{pages}{124601}.
%Type = Article
\bibitem[{Spindler et~al.(2016)Spindler, Ehrig, König, Nowak, Piliarik, Stein,
  Taylor, Garanger, Lecommandoux, Alves, and Sandoghdar}]{Spindler_2016}
\bibinfo{author}{S.~Spindler}, \bibinfo{author}{J.~Ehrig},
  \bibinfo{author}{K.~König}, \bibinfo{author}{T.~Nowak},
  \bibinfo{author}{M.~Piliarik}, \bibinfo{author}{H.~E. Stein},
  \bibinfo{author}{R.~W. Taylor}, \bibinfo{author}{E.~Garanger},
  \bibinfo{author}{S.~Lecommandoux}, \bibinfo{author}{I.~D. Alves},
  \bibinfo{author}{V.~Sandoghdar},
\newblock \bibinfo{title}{Visualization of lipids and proteins at high spatial
  and temporal resolution via interferometric scattering ({iSCAT}) microscopy},
\newblock \bibinfo{journal}{J. Phys. D} \bibinfo{volume}{49}
  (\bibinfo{year}{2016}) \bibinfo{pages}{274002}.
%Type = Article
\bibitem[{Taylor et~al.(2019)Taylor, Mahmoodabadi, Rauschenberger, Giessl,
  Schambony, and Sandoghdar}]{Taylor_2019a}
\bibinfo{author}{R.~W. Taylor}, \bibinfo{author}{R.~G. Mahmoodabadi},
  \bibinfo{author}{V.~Rauschenberger}, \bibinfo{author}{A.~Giessl},
  \bibinfo{author}{A.~Schambony}, \bibinfo{author}{V.~Sandoghdar},
\newblock \bibinfo{title}{Interferometric scattering microscopy reveals
  microsecond nanoscopic protein motion on a live cell membrane},
\newblock \bibinfo{journal}{Nature Photonics} \bibinfo{volume}{13}
  (\bibinfo{year}{2019}) \bibinfo{pages}{480--487}.
%Type = Article
\bibitem[{Shen et~al.(2017)Shen, Tauzin, Baiyasi, Wang, Moringo, Shuang, and
  Landes}]{Shen_2017}
\bibinfo{author}{H.~Shen}, \bibinfo{author}{L.~J. Tauzin},
  \bibinfo{author}{R.~Baiyasi}, \bibinfo{author}{W.~Wang},
  \bibinfo{author}{N.~Moringo}, \bibinfo{author}{B.~Shuang},
  \bibinfo{author}{C.~F. Landes},
\newblock \bibinfo{title}{Single particle tracking: From theory to biophysical
  applications},
\newblock \bibinfo{journal}{Chem. Rev.} \bibinfo{volume}{117}
  (\bibinfo{year}{2017}) \bibinfo{pages}{7331--7376}.
%Type = Article
\bibitem[{Araque et~al.(2015)Araque, Yadav, Shadeck, Maroncelli, and
  Margulis}]{Araque_2015}
\bibinfo{author}{J.~C. Araque}, \bibinfo{author}{S.~K. Yadav},
  \bibinfo{author}{M.~Shadeck}, \bibinfo{author}{M.~Maroncelli},
  \bibinfo{author}{C.~J. Margulis},
\newblock \bibinfo{title}{How is diffusion of neutral and charged tracers
  related to the structure and dynamics of a room-temperature ionic liquid?
  large deviations from stokes–einstein behavior explained},
\newblock \bibinfo{journal}{J. Phys. Chem. B} \bibinfo{volume}{119}
  (\bibinfo{year}{2015}) \bibinfo{pages}{7015--7029}.
%Type = Article
\bibitem[{Kirstein et~al.(2007)Kirstein, Platschek, Jung, Brown, Bein, and
  Bräuchle}]{Kirstein_2007}
\bibinfo{author}{J.~Kirstein}, \bibinfo{author}{B.~Platschek},
  \bibinfo{author}{C.~Jung}, \bibinfo{author}{R.~Brown},
  \bibinfo{author}{T.~Bein}, \bibinfo{author}{C.~Bräuchle},
\newblock \bibinfo{title}{Exploration of nanostructured channel systems with
  single-molecule probes},
\newblock \bibinfo{journal}{Nat. Mater.} \bibinfo{volume}{6}
  (\bibinfo{year}{2007}) \bibinfo{pages}{303--310}.
%Type = Article
\bibitem[{Higgins et~al.(2015)Higgins, Park, Tran-Ba, and Ito}]{Higgins_2015}
\bibinfo{author}{D.~A. Higgins}, \bibinfo{author}{S.~C. Park},
  \bibinfo{author}{K.-H. Tran-Ba}, \bibinfo{author}{T.~Ito},
\newblock \bibinfo{title}{Single-molecule investigations of morphology and mass
  transport dynamics in nanostructured materials},
\newblock \bibinfo{journal}{Annu. Rev. Anal. Chem.} \bibinfo{volume}{8}
  (\bibinfo{year}{2015}) \bibinfo{pages}{193--216}.
%Type = Article
\bibitem[{Grier(1997)}]{Grier_1997}
\bibinfo{author}{D.~G. Grier},
\newblock \bibinfo{title}{Optical tweezers in colloid and interface science},
\newblock \bibinfo{journal}{Curr. Opin. Colloid Interface Sci.}
  \bibinfo{volume}{2} (\bibinfo{year}{1997}) \bibinfo{pages}{264--270}.
%Type = Article
\bibitem[{Neuman and Block(2004)}]{Neuman_2004}
\bibinfo{author}{K.~C. Neuman}, \bibinfo{author}{S.~M. Block},
\newblock \bibinfo{title}{Optical trapping},
\newblock \bibinfo{journal}{Rev. Sci. Instrum.} \bibinfo{volume}{75}
  (\bibinfo{year}{2004}) \bibinfo{pages}{2787--2809}.
%Type = Article
\bibitem[{Maragò et~al.(2013)Maragò, Jones, Gucciardi, Volpe, and
  Ferrari}]{Marago_2013}
\bibinfo{author}{O.~M. Maragò}, \bibinfo{author}{P.~H. Jones},
  \bibinfo{author}{P.~G. Gucciardi}, \bibinfo{author}{G.~Volpe},
  \bibinfo{author}{A.~C. Ferrari},
\newblock \bibinfo{title}{Optical trapping and manipulation of nanostructures},
\newblock \bibinfo{journal}{Nat. Nanotechnol.} \bibinfo{volume}{8}
  (\bibinfo{year}{2013}) \bibinfo{pages}{807}.
%Type = Article
\bibitem[{Bowman and Padgett(2013)}]{Bowman_2013}
\bibinfo{author}{R.~W. Bowman}, \bibinfo{author}{M.~J. Padgett},
\newblock \bibinfo{title}{Optical trapping and binding},
\newblock \bibinfo{journal}{Rep. Prog. Phys.} \bibinfo{volume}{76}
  (\bibinfo{year}{2013}) \bibinfo{pages}{026401}.
%Type = Book
\bibitem[{Philip H.~Jones(2015)}]{Jones_2015}
\bibinfo{author}{G.~V. Philip H.~Jones, Onofrio M.~Maragò},
  \bibinfo{title}{Optical tweezers : principles and applications},
  \bibinfo{publisher}{Cambridge University Press},
  \bibinfo{address}{Cambridge}, \bibinfo{year}{2015}.
%Type = Article
\bibitem[{Bustamante et~al.(2021)Bustamante, Chemla, Liu, and
  Wang}]{Bustamante_2021}
\bibinfo{author}{C.~J. Bustamante}, \bibinfo{author}{Y.~R. Chemla},
  \bibinfo{author}{S.~Liu}, \bibinfo{author}{M.~D. Wang},
\newblock \bibinfo{title}{Optical tweezers in single-molecule biophysics},
\newblock \bibinfo{journal}{Nat. Rev. Methods Primers} \bibinfo{volume}{1}
  (\bibinfo{year}{2021}) \bibinfo{pages}{25}.
%Type = Article
\bibitem[{Ashkin et~al.(1986)Ashkin, Dziedzic, Bjorkholm, and
  Chu}]{Ashkin_1986}
\bibinfo{author}{A.~Ashkin}, \bibinfo{author}{J.~M. Dziedzic},
  \bibinfo{author}{J.~E. Bjorkholm}, \bibinfo{author}{S.~Chu},
\newblock \bibinfo{title}{Observation of a single-beam gradient force optical
  trap for dielectric particles},
\newblock \bibinfo{journal}{Opt. Lett.} \bibinfo{volume}{11}
  (\bibinfo{year}{1986}) \bibinfo{pages}{288--290}.
%Type = Article
\bibitem[{Ashkin and Dziedzic(1987)}]{Ashkin_1987}
\bibinfo{author}{A.~Ashkin}, \bibinfo{author}{J.~M. Dziedzic},
\newblock \bibinfo{title}{Optical trapping and manipulation of viruses and
  bacteria},
\newblock \bibinfo{journal}{Science} \bibinfo{volume}{235}
  (\bibinfo{year}{1987}) \bibinfo{pages}{1517--1520}.
%Type = Article
\bibitem[{Ashkin et~al.(1987)Ashkin, Dziedzic, and Yamane}]{Ashkin_1987a}
\bibinfo{author}{A.~Ashkin}, \bibinfo{author}{J.~M. Dziedzic},
  \bibinfo{author}{T.~Yamane},
\newblock \bibinfo{title}{Optical trapping and manipulation of single cells
  using infrared laser beams},
\newblock \bibinfo{journal}{Nature} \bibinfo{volume}{330}
  (\bibinfo{year}{1987}) \bibinfo{pages}{769--771}.
%Type = Article
\bibitem[{Svoboda et~al.(1993)Svoboda, Schmidt, Schnapp, and
  Block}]{Svoboda_1993}
\bibinfo{author}{K.~Svoboda}, \bibinfo{author}{C.~F. Schmidt},
  \bibinfo{author}{B.~J. Schnapp}, \bibinfo{author}{S.~M. Block},
\newblock \bibinfo{title}{Direct observation of kinesin stepping by optical
  trapping interferometry},
\newblock \bibinfo{journal}{Nature} \bibinfo{volume}{365}
  (\bibinfo{year}{1993}) \bibinfo{pages}{721--727}.
%Type = Article
\bibitem[{Wang et~al.(1997)Wang, Yin, Landick, Gelles, and Block}]{Wang_1997}
\bibinfo{author}{M.~Wang}, \bibinfo{author}{H.~Yin},
  \bibinfo{author}{R.~Landick}, \bibinfo{author}{J.~Gelles},
  \bibinfo{author}{S.~Block},
\newblock \bibinfo{title}{Stretching dna with optical tweezers},
\newblock \bibinfo{journal}{Biophys. J.} \bibinfo{volume}{72}
  (\bibinfo{year}{1997}) \bibinfo{pages}{1335--1346}.
%Type = Article
\bibitem[{Lang et~al.(2004)Lang, Fordyce, Engh, Neuman, and Block}]{Lang_2004}
\bibinfo{author}{M.~J. Lang}, \bibinfo{author}{P.~M. Fordyce},
  \bibinfo{author}{A.~M. Engh}, \bibinfo{author}{K.~C. Neuman},
  \bibinfo{author}{S.~M. Block},
\newblock \bibinfo{title}{Simultaneous, coincident optical trapping and
  single-molecule fluorescence},
\newblock \bibinfo{journal}{Nat. Methods} \bibinfo{volume}{1}
  (\bibinfo{year}{2004}) \bibinfo{pages}{133--139}.
%Type = Article
\bibitem[{Comstock et~al.(2011)Comstock, Ha, and Chemla}]{Comstock_2011}
\bibinfo{author}{M.~J. Comstock}, \bibinfo{author}{T.~Ha},
  \bibinfo{author}{Y.~R. Chemla},
\newblock \bibinfo{title}{Ultrahigh-resolution optical trap with
  single-fluorophore sensitivity},
\newblock \bibinfo{journal}{Nat. Methods} \bibinfo{volume}{8}
  (\bibinfo{year}{2011}) \bibinfo{pages}{335--340}.
%Type = Article
\bibitem[{Yao et~al.(2009)Yao, Tassieri, Padgett, and Cooper}]{Yao_2009}
\bibinfo{author}{A.~Yao}, \bibinfo{author}{M.~Tassieri},
  \bibinfo{author}{M.~Padgett}, \bibinfo{author}{J.~Cooper},
\newblock \bibinfo{title}{Microrheology with optical tweezers},
\newblock \bibinfo{journal}{Lab Chip} \bibinfo{volume}{9}
  (\bibinfo{year}{2009}) \bibinfo{pages}{2568--2575}.
%Type = Article
\bibitem[{Polin et~al.(2006)Polin, Grier, and Quake}]{Polin_2006}
\bibinfo{author}{M.~Polin}, \bibinfo{author}{D.~G. Grier},
  \bibinfo{author}{S.~R. Quake},
\newblock \bibinfo{title}{Anomalous vibrational dispersion in holographically
  trapped colloidal arrays},
\newblock \bibinfo{journal}{Phys. Rev. Lett.} \bibinfo{volume}{96}
  (\bibinfo{year}{2006}) \bibinfo{pages}{088101}.
%Type = Article
\bibitem[{Walder et~al.(2011)Walder, Nelson, and Schwartz}]{Walder_2011}
\bibinfo{author}{R.~Walder}, \bibinfo{author}{N.~Nelson},
  \bibinfo{author}{D.~K. Schwartz},
\newblock \bibinfo{title}{Single molecule observations of desorption-mediated
  diffusion at the solid-liquid interface},
\newblock \bibinfo{journal}{Phys. Rev. Lett.} \bibinfo{volume}{107}
  (\bibinfo{year}{2011}) \bibinfo{pages}{156102}.
%Type = Article
\bibitem[{Skaug et~al.(2014)Skaug, Mabry, and Schwartz}]{Skaug_2014}
\bibinfo{author}{M.~J. Skaug}, \bibinfo{author}{J.~N. Mabry},
  \bibinfo{author}{D.~K. Schwartz},
\newblock \bibinfo{title}{Single-molecule tracking of polymer surface
  diffusion},
\newblock \bibinfo{journal}{J. Am. Chem. Soc.} \bibinfo{volume}{136}
  (\bibinfo{year}{2014}) \bibinfo{pages}{1327--1332}.
%Type = Article
\bibitem[{Dong et~al.(2018)Dong, Pei, Zhao, Goh, Qi, Xiao, Chen, Huang, and
  Fang}]{Dong_2018}
\bibinfo{author}{B.~Dong}, \bibinfo{author}{Y.~Pei}, \bibinfo{author}{F.~Zhao},
  \bibinfo{author}{T.~W. Goh}, \bibinfo{author}{Z.~Qi},
  \bibinfo{author}{C.~Xiao}, \bibinfo{author}{K.~Chen},
  \bibinfo{author}{W.~Huang}, \bibinfo{author}{N.~Fang},
\newblock \bibinfo{title}{In situ quantitative single-molecule study of dynamic
  catalytic processes in nanoconfinement},
\newblock \bibinfo{journal}{Nat. Catal.} \bibinfo{volume}{1}
  (\bibinfo{year}{2018}) \bibinfo{pages}{135--140}.
%Type = Article
\bibitem[{Zhong and Wang(2020)}]{Zhong_2020}
\bibinfo{author}{Y.~Zhong}, \bibinfo{author}{G.~Wang},
\newblock \bibinfo{title}{Three-dimensional single particle tracking and its
  applications in confined environments},
\newblock \bibinfo{journal}{Annu. Rev. Anal. Chem.} \bibinfo{volume}{13}
  (\bibinfo{year}{2020}) \bibinfo{pages}{381--403}.
%Type = Book
\bibitem[{Atkins and De~Paula(2014)}]{Atkins_2014a}
\bibinfo{author}{P.~Atkins}, \bibinfo{author}{J.~De~Paula},
  \bibinfo{title}{Atkins’ Physical Chemistry}, \bibinfo{publisher}{Oxford
  University Press}, \bibinfo{address}{Oxford, UK}, \bibinfo{year}{2014}.
%Type = Article
\bibitem[{Yuan et~al.(2017)Yuan, Lin, Gu, Panwar, Tjin, Song, Qu, and
  Yong}]{Yuan_2017b}
\bibinfo{author}{Y.~Yuan}, \bibinfo{author}{Y.~Lin}, \bibinfo{author}{B.~Gu},
  \bibinfo{author}{N.~Panwar}, \bibinfo{author}{S.~C. Tjin},
  \bibinfo{author}{J.~Song}, \bibinfo{author}{J.~Qu}, \bibinfo{author}{K.-T.
  Yong},
\newblock \bibinfo{title}{Optical trapping-assisted sers platform for chemical
  and biosensing applications: Design perspectives},
\newblock \bibinfo{journal}{Coord. Chem. Rev.} \bibinfo{volume}{339}
  (\bibinfo{year}{2017}) \bibinfo{pages}{138--152}.
%Type = Article
\bibitem[{Oyamada et~al.(2022)Oyamada, Minamimoto, and
  Murakoshi}]{Oyamada_2022}
\bibinfo{author}{N.~Oyamada}, \bibinfo{author}{H.~Minamimoto},
  \bibinfo{author}{K.~Murakoshi},
\newblock \bibinfo{title}{Room-temperature molecular manipulation via plasmonic
  trapping at electrified interfaces},
\newblock \bibinfo{journal}{J. Am. Chem. Soc.}  (\bibinfo{year}{2022}).
%Type = Article
\bibitem[{Tsuboi et~al.(2021)Tsuboi, Naka, Yamanishi, Nagai, Yuyama, Shoji,
  Ohtani, Tamura, Iida, Kameyama, and Torimoto}]{Tsuboi_2021}
\bibinfo{author}{Y.~Tsuboi}, \bibinfo{author}{S.~Naka},
  \bibinfo{author}{D.~Yamanishi}, \bibinfo{author}{T.~Nagai},
  \bibinfo{author}{K.-i. Yuyama}, \bibinfo{author}{T.~Shoji},
  \bibinfo{author}{B.~Ohtani}, \bibinfo{author}{M.~Tamura},
  \bibinfo{author}{T.~Iida}, \bibinfo{author}{T.~Kameyama},
  \bibinfo{author}{T.~Torimoto},
\newblock \bibinfo{title}{Optical trapping of nanocrystals at oil/water
  interfaces: Implications for photocatalysis},
\newblock \bibinfo{journal}{ACS Appl. Nano Mater.} \bibinfo{volume}{4}
  (\bibinfo{year}{2021}) \bibinfo{pages}{11743--11752}.
%Type = Article
\bibitem[{Ducker and Wanless(1999)}]{Ducker_1999}
\bibinfo{author}{W.~A. Ducker}, \bibinfo{author}{E.~J. Wanless},
\newblock \bibinfo{title}{Adsorption of hexadecyltrimethylammonium bromide to
  mica: Nanometer-scale study of binding-site competition effects},
\newblock \bibinfo{journal}{Langmuir} \bibinfo{volume}{15}
  (\bibinfo{year}{1999}) \bibinfo{pages}{160--168}.
%Type = Article
\bibitem[{Tyrode et~al.(2008)Tyrode, Rutland, and Bain}]{Tyrode_2008}
\bibinfo{author}{E.~Tyrode}, \bibinfo{author}{M.~W. Rutland},
  \bibinfo{author}{C.~D. Bain},
\newblock \bibinfo{title}{Adsorption of ctab on hydrophilic silica studied by
  linear and nonlinear optical spectroscopy},
\newblock \bibinfo{journal}{J. Am. Chem. Soc.} \bibinfo{volume}{130}
  (\bibinfo{year}{2008}) \bibinfo{pages}{17434--17445}.
%Type = Article
\bibitem[{Evans and Needham(1987)}]{Evans_1987}
\bibinfo{author}{E.~Evans}, \bibinfo{author}{D.~Needham},
\newblock \bibinfo{title}{Physical properties of surfactant bilayer membranes:
  thermal transitions, elasticity, rigidity, cohesion and colloidal
  interactions},
\newblock \bibinfo{journal}{J. Phys. Chem.} \bibinfo{volume}{91}
  (\bibinfo{year}{1987}) \bibinfo{pages}{4219--4228}.
%Type = Article
\bibitem[{Li et~al.(2020)Li, Wang, Gu, Chen, Zhang, and Hu}]{Li_2020}
\bibinfo{author}{R.~Li}, \bibinfo{author}{Z.~Wang}, \bibinfo{author}{X.~Gu},
  \bibinfo{author}{C.~Chen}, \bibinfo{author}{Y.~Zhang},
  \bibinfo{author}{D.~Hu},
\newblock \bibinfo{title}{Study on the assembly structure variation of
  cetyltrimethylammonium bromide on the surface of gold nanoparticles},
\newblock \bibinfo{journal}{ACS Omega} \bibinfo{volume}{5}
  (\bibinfo{year}{2020}) \bibinfo{pages}{4943--4952}.
%Type = Article
\bibitem[{Kottmann et~al.(2001)Kottmann, Martin, Smith, and
  Schultz}]{Kottmann_2001e}
\bibinfo{author}{J.~P. Kottmann}, \bibinfo{author}{O.~J.~F. Martin},
  \bibinfo{author}{D.~R. Smith}, \bibinfo{author}{S.~Schultz},
\newblock \bibinfo{title}{Dramatic localized electromagnetic enhancement in
  plasmon resonant nanowires},
\newblock \bibinfo{journal}{Chem. Phys. Lett.} \bibinfo{volume}{341}
  (\bibinfo{year}{2001}) \bibinfo{pages}{1--6}.
%Type = Article
\bibitem[{Taylor and Sandoghdar(2019)}]{Taylor_2019}
\bibinfo{author}{R.~W. Taylor}, \bibinfo{author}{V.~Sandoghdar},
\newblock \bibinfo{title}{Interferometric scattering microscopy: Seeing single
  nanoparticles and molecules via rayleigh scattering},
\newblock \bibinfo{journal}{Nano Lett.} \bibinfo{volume}{19}
  (\bibinfo{year}{2019}) \bibinfo{pages}{4827--4835}.
%Type = Article
\bibitem[{Alam et~al.(2014)Alam, Howlader, and Deen}]{Alam_2014a}
\bibinfo{author}{A.~U. Alam}, \bibinfo{author}{M.~M.~R. Howlader},
  \bibinfo{author}{M.~J. Deen},
\newblock \bibinfo{title}{The effects of oxygen plasma and humidity on surface
  roughness, water contact angle and hardness of silicon, silicon dioxide and
  glass},
\newblock \bibinfo{journal}{J. Micromech. Microeng.} \bibinfo{volume}{24}
  (\bibinfo{year}{2014}) \bibinfo{pages}{035010}.
%Type = Misc
\bibitem[{Allan et~al.(2021)Allan, Caswell, Keim, van~der Wel, and
  Verweij}]{Allan_2021}
\bibinfo{author}{D.~B. Allan}, \bibinfo{author}{T.~Caswell},
  \bibinfo{author}{N.~C. Keim}, \bibinfo{author}{C.~M. van~der Wel},
  \bibinfo{author}{R.~W. Verweij}, \bibinfo{title}{soft-matter/trackpy: Trackpy
  v0.5.0}, \bibinfo{year}{2021}. \URLprefix
  \url{https://doi.org/10.5281/zenodo.4682814}.
  \DOIprefix\doi{10.5281/zenodo.4682814}.
%Type = Article
\bibitem[{Crocker and Grier(1996)}]{Crocker_1996}
\bibinfo{author}{J.~C. Crocker}, \bibinfo{author}{D.~G. Grier},
\newblock \bibinfo{title}{Methods of digital video microscopy for colloidal
  studies},
\newblock \bibinfo{journal}{J. Colloid Interface Sci.} \bibinfo{volume}{179}
  (\bibinfo{year}{1996}) \bibinfo{pages}{298--310}.
%Type = Article
\bibitem[{Kim and Martin(2022{\natexlab{a}})}]{Kim_2021e}
\bibinfo{author}{J.~Kim}, \bibinfo{author}{O.~J.~F. Martin},
\newblock \bibinfo{title}{Surfactants control optical trapping near a glass
  wall},
\newblock \bibinfo{journal}{J. Phys. Chem. C} \bibinfo{volume}{126}
  (\bibinfo{year}{2022}{\natexlab{a}}) \bibinfo{pages}{378--386}.
%Type = Misc
\bibitem[{Kim and Martin(2022{\natexlab{b}})}]{Kim_2022}
\bibinfo{author}{J.~Kim}, \bibinfo{author}{O.~J.~F. Martin},
  \bibinfo{title}{{Dataset for the manuscript: Probing surfactant bilayer
  interactions by tracking optically trapped single nanoparticles}},
  \bibinfo{year}{2022}{\natexlab{b}}. \URLprefix
  \url{https://doi.org/10.5281/zenodo.6015007}.
  \DOIprefix\doi{10.5281/zenodo.6015007}.
%Type = Article
\bibitem[{Atkin et~al.(2003)Atkin, Craig, Wanless, and Biggs}]{Atkin_2003}
\bibinfo{author}{R.~Atkin}, \bibinfo{author}{V.~Craig},
  \bibinfo{author}{E.~Wanless}, \bibinfo{author}{S.~Biggs},
\newblock \bibinfo{title}{Mechanism of cationic surfactant adsorption at the
  solid–aqueous interface},
\newblock \bibinfo{journal}{Adv. Colloid Interface Sci.} \bibinfo{volume}{103}
  (\bibinfo{year}{2003}) \bibinfo{pages}{219--304}.
%Type = Article
\bibitem[{Goloub et~al.(1996)Goloub, Koopal, Bijsterbosch, and
  Sidorova}]{Goloub_1996}
\bibinfo{author}{T.~P. Goloub}, \bibinfo{author}{L.~K. Koopal},
  \bibinfo{author}{B.~H. Bijsterbosch}, \bibinfo{author}{M.~P. Sidorova},
\newblock \bibinfo{title}{Adsorption of cationic surfactants on silica. surface
  charge effects},
\newblock \bibinfo{journal}{Langmuir} \bibinfo{volume}{12}
  (\bibinfo{year}{1996}) \bibinfo{pages}{3188--3194}.
%Type = Article
\bibitem[{Goloub and Koopal(1997)}]{Goloub_1997}
\bibinfo{author}{T.~P. Goloub}, \bibinfo{author}{L.~K. Koopal},
\newblock \bibinfo{title}{Adsorption of cationic surfactants on silica.
  comparison of experiment and theory},
\newblock \bibinfo{journal}{Langmuir} \bibinfo{volume}{13}
  (\bibinfo{year}{1997}) \bibinfo{pages}{673--681}.
%Type = Article
\bibitem[{Lamont and Ducker(1998)}]{Lamont_1998}
\bibinfo{author}{R.~E. Lamont}, \bibinfo{author}{W.~A. Ducker},
\newblock \bibinfo{title}{Surface-induced transformations for surfactant
  aggregates},
\newblock \bibinfo{journal}{J. Am. Chem. Soc.} \bibinfo{volume}{120}
  (\bibinfo{year}{1998}) \bibinfo{pages}{7602--7607}.
%Type = Article
\bibitem[{Chorro et~al.(1999)Chorro, Chorro, Dolladille, Partyka, and
  Zana}]{Chorro_1999}
\bibinfo{author}{M.~Chorro}, \bibinfo{author}{C.~Chorro},
  \bibinfo{author}{O.~Dolladille}, \bibinfo{author}{S.~Partyka},
  \bibinfo{author}{R.~Zana},
\newblock \bibinfo{title}{Adsorption mechanism of conventional and dimeric
  cationic surfactants on silica surface: Effect of the state of the surface},
\newblock \bibinfo{journal}{J. Colloid Interface Sci.} \bibinfo{volume}{210}
  (\bibinfo{year}{1999}) \bibinfo{pages}{134--143}.
%Type = Article
\bibitem[{Grant et~al.(2000)Grant, Ederth, and Tiberg}]{Grant_2000}
\bibinfo{author}{L.~M. Grant}, \bibinfo{author}{T.~Ederth},
  \bibinfo{author}{F.~Tiberg},
\newblock \bibinfo{title}{Influence of surface hydrophobicity on the layer
  properties of adsorbed nonionic surfactants},
\newblock \bibinfo{journal}{Langmuir} \bibinfo{volume}{16}
  (\bibinfo{year}{2000}) \bibinfo{pages}{2285--2291}.
%Type = Article
\bibitem[{Behrens and Grier(2001)}]{Behrens_2001}
\bibinfo{author}{S.~H. Behrens}, \bibinfo{author}{D.~G. Grier},
\newblock \bibinfo{title}{The charge of glass and silica surfaces},
\newblock \bibinfo{journal}{J. Chem. Phys.} \bibinfo{volume}{115}
  (\bibinfo{year}{2001}) \bibinfo{pages}{6716--6721}.
%Type = Article
\bibitem[{Helm et~al.(1992)Helm, Israelachvili, and McGuiggan}]{Helm_1992}
\bibinfo{author}{C.~A. Helm}, \bibinfo{author}{J.~N. Israelachvili},
  \bibinfo{author}{P.~M. McGuiggan},
\newblock \bibinfo{title}{Role of hydrophobic forces in bilayer adhesion and
  fusion},
\newblock \bibinfo{journal}{Biochem.} \bibinfo{volume}{31}
  (\bibinfo{year}{1992}) \bibinfo{pages}{1794--1805}.
%Type = Article
\bibitem[{Petrache et~al.(1998)Petrache, Gouliaev, Tristram-Nagle, Zhang,
  Suter, and Nagle}]{Petrache_1998}
\bibinfo{author}{H.~I. Petrache}, \bibinfo{author}{N.~Gouliaev},
  \bibinfo{author}{S.~Tristram-Nagle}, \bibinfo{author}{R.~Zhang},
  \bibinfo{author}{R.~M. Suter}, \bibinfo{author}{J.~F. Nagle},
\newblock \bibinfo{title}{Interbilayer interactions from high-resolution x-ray
  scattering},
\newblock \bibinfo{journal}{Phys. Rev. E} \bibinfo{volume}{57}
  (\bibinfo{year}{1998}) \bibinfo{pages}{7014--7024}.
%Type = Inbook
\bibitem[{Atkins and De~Paula(2014)}]{Atkins_2014}
\bibinfo{author}{P.~Atkins}, \bibinfo{author}{J.~De~Paula},
  \bibinfo{title}{Atkins’ Physical Chemistry}, \bibinfo{publisher}{Oxford
  University Press}, \bibinfo{address}{Oxford, UK}, \bibinfo{year}{2014}, pp.
  \bibinfo{pages}{628--629}.
%Type = Article
\bibitem[{N\o{}rrelykke and Flyvbjerg(2011)}]{Norrelykke_2011}
\bibinfo{author}{S.~F. N\o{}rrelykke}, \bibinfo{author}{H.~Flyvbjerg},
\newblock \bibinfo{title}{Harmonic oscillator in heat bath: Exact simulation of
  time-lapse-recorded data and exact analytical benchmark statistics},
\newblock \bibinfo{journal}{Phys. Rev. E} \bibinfo{volume}{83}
  (\bibinfo{year}{2011}) \bibinfo{pages}{041103}.
%Type = Article
\bibitem[{Purcell(1977)}]{Purcell_1977}
\bibinfo{author}{E.~M. Purcell},
\newblock \bibinfo{title}{Life at low reynolds number},
\newblock \bibinfo{journal}{Am. J. Phys.} \bibinfo{volume}{45}
  (\bibinfo{year}{1977}) \bibinfo{pages}{3--11}.
%Type = Book
\bibitem[{Happel and Brenner(2012)}]{Happel_2012}
\bibinfo{author}{J.~Happel}, \bibinfo{author}{H.~Brenner}, \bibinfo{title}{Low
  Reynolds number hydrodynamics: with special applications to particulate
  media}, volume~\bibinfo{volume}{1}, \bibinfo{publisher}{Springer, Dordrecht},
  \bibinfo{year}{2012}.
%Type = Article
\bibitem[{Svoboda and Block(1994)}]{Svoboda_1994a}
\bibinfo{author}{K.~Svoboda}, \bibinfo{author}{S.~M. Block},
\newblock \bibinfo{title}{Biological applications of optical forces},
\newblock \bibinfo{journal}{Annu. Rev. Biophys. Biomol. Struct.}
  \bibinfo{volume}{23} (\bibinfo{year}{1994}) \bibinfo{pages}{247--285}.
%Type = Article
\bibitem[{Neusius et~al.(2009)Neusius, Sokolov, and Smith}]{Neusius_2009}
\bibinfo{author}{T.~Neusius}, \bibinfo{author}{I.~M. Sokolov},
  \bibinfo{author}{J.~C. Smith},
\newblock \bibinfo{title}{Subdiffusion in time-averaged, confined random
  walks},
\newblock \bibinfo{journal}{Phys. Rev. E} \bibinfo{volume}{80}
  (\bibinfo{year}{2009}) \bibinfo{pages}{011109}.
%Type = Article
\bibitem[{Jeon and Metzler(2010)}]{Jeon_2010}
\bibinfo{author}{J.-H. Jeon}, \bibinfo{author}{R.~Metzler},
\newblock \bibinfo{title}{Fractional brownian motion and motion governed by the
  fractional langevin equation in confined geometries},
\newblock \bibinfo{journal}{Phys. Rev. E} \bibinfo{volume}{81}
  (\bibinfo{year}{2010}) \bibinfo{pages}{021103}.
%Type = Article
\bibitem[{Toli\'{c}-N\o{}rrelykke et~al.(2004)Toli\'{c}-N\o{}rrelykke,
  Munteanu, Thon, Oddershede, and Berg-S\o{}rensen}]{Toli_2004}
\bibinfo{author}{I.~M. Toli\'{c}-N\o{}rrelykke}, \bibinfo{author}{E.-L.
  Munteanu}, \bibinfo{author}{G.~Thon}, \bibinfo{author}{L.~Oddershede},
  \bibinfo{author}{K.~Berg-S\o{}rensen},
\newblock \bibinfo{title}{Anomalous diffusion in living yeast cells},
\newblock \bibinfo{journal}{Phys. Rev. Lett.} \bibinfo{volume}{93}
  (\bibinfo{year}{2004}) \bibinfo{pages}{078102}.
%Type = Article
\bibitem[{Wachsmuth et~al.(2000)Wachsmuth, Waldeck, and
  Langowski}]{Wachsmuth_2000}
\bibinfo{author}{M.~Wachsmuth}, \bibinfo{author}{W.~Waldeck},
  \bibinfo{author}{J.~Langowski},
\newblock \bibinfo{title}{Anomalous diffusion of fluorescent probes inside
  living cell nuclei investigated by spatially-resolved fluorescence
  correlation spectroscopy},
\newblock \bibinfo{journal}{J. Mol. Biol.} \bibinfo{volume}{298}
  (\bibinfo{year}{2000}) \bibinfo{pages}{677--689}.
%Type = Article
\bibitem[{Jeon et~al.(2013)Jeon, Leijnse, Oddershede, and Metzler}]{Jeon_2013}
\bibinfo{author}{J.-H. Jeon}, \bibinfo{author}{N.~Leijnse},
  \bibinfo{author}{L.~B. Oddershede}, \bibinfo{author}{R.~Metzler},
\newblock \bibinfo{title}{Anomalous diffusion and power-law relaxation of the
  time averaged mean squared displacement in worm-like micellar solutions},
\newblock \bibinfo{journal}{New J. Phys.} \bibinfo{volume}{15}
  (\bibinfo{year}{2013}) \bibinfo{pages}{045011}.

\end{thebibliography}

%% else use the following coding to input the bibitems directly in the
%% TeX file.

% \begin{thebibliography}{00}

% %% \bibitem{label}
% %% Text of bibliographic item

% \bibitem{}

% \end{thebibliography}
\end{document}